\documentclass[12pt]{article}

\usepackage{hyperref}

\usepackage{amsmath}
\usepackage{amssymb}
\usepackage{xspace}
\usepackage{graphicx, subfig}               
\usepackage{setspace}
\usepackage[authoryear, longnamesfirst, sort&compress]{natbib}

\usepackage{color}

\newcommand{\ac}{\color{red} }  
\newcommand{\ca}{ \color{black}} 
\newcommand{\cutac}[1]{\ac [cut] \ca} 

\newcommand{\by}{\mathbf{y}}
\newcommand{\bY}{\mathbf{Y}}
\newcommand{\logten}{\log_{10}}

\renewenvironment{abstract}
 {\small
  \begin{center}
  \bfseries \abstractname\vspace{-.5em}\vspace{0pt}
  \end{center}
  \list{}{%
    \setlength{\leftmargin}{3mm}
    \setlength{\rightmargin}{\leftmargin}%
  }%
  \item\relax}
 {\endlist}

\usepackage[bindingoffset=0.0in,%
            left=1in,right=1in,top=1in,bottom=1in,%
            footskip=.25in]{geometry}

\date{12MAR2020}

\newcommand{\papertitle}{Systematic statistical analysis of microbial data from dilution series}

\newcommand{\blind}{1}
\newcommand{\acknow}{JAC is partially funded by CONACYT CB-2016-01-284451, RDECOMM and ONRG grants. AEP is partially funded by NSF grant 1516951.  The authors gratefully acknowledge the industrial associates of the CBE, Lindsey Lorenz and  Professor Emeritus Martin Hamilton.}

\begin{document}

\def\spacingset#1{\renewcommand{\baselinestretch}{#1}\small\normalsize}
\spacingset{1}


\if1\blind
{
  \title{\textbf{\papertitle}}
    \author{J Andr\'es Christen\thanks{J Andr\'es Christen, Centro de Investigaci\'on en Matem\'aticas (CIMAT-CONACYT), Jalisco S/N, Valenciana, Guanajuato, GTO, 36023, MEXICO, email:\textit{jac at cimat.mx}.} \thanks{Corresponding author.}  \and Albert E. Parker\thanks{Center for Biofilm Engineering (CBE) and Department of Mathematical Sciences, Montana State University, Bozeman, MT, USA, email:  \textit{parker at math.montana.edu}.}}
  \maketitle
} \fi

\if0\blind
{
  \bigskip
  \bigskip
  \bigskip
  \begin{center}
    {\LARGE\bf \papertitle}
\end{center}
  \medskip
  \centerline{Blinded authors}
} \fi

\begin{abstract}
In microbial studies, samples are often treated under different experimental conditions and then tested for microbial survival.
A technique, dating back to the 1880's, consists of diluting the samples several times and incubating
each dilution to verify the existence of microbial Colony Forming Units or CFU's, seen by the naked   eye.
The main problem in the dilution series data analysis is
the uncertainty quantification of the simple point estimate of the original number of CFU's in the sample (i.e., at dilution zero).  Common approaches
such as log-normal or Poisson models do not seem to handle well extreme cases with low or high counts, among other issues.  We build a novel binomial model, based on the actual design of the experimental procedure including the dilution series.
For repetitions we construct a hierarchical model for experimental results from a single lab and in turn a higher hierarchy
for inter-lab analyses.  Results seem promising, with a systematic treatment of all data cases, including zeros,
censored data, repetitions, intra and inter-laboratory studies.  Using a Bayesian approach, a robust and efficient
MCMC method is used to analyze several real data sets.
\end{abstract}

\noindent%
{\it Keywords:}  Dilution experiments; binomial likelihood; Bayesian inference; Hierarchical models; MCMC.
\vfill


\spacingset{1.4} 


\section{Introduction}\label{sec:intro}

\begin{figure}
\begin{center}
\begin{tabular}{c c c}
\includegraphics[scale=0.048]{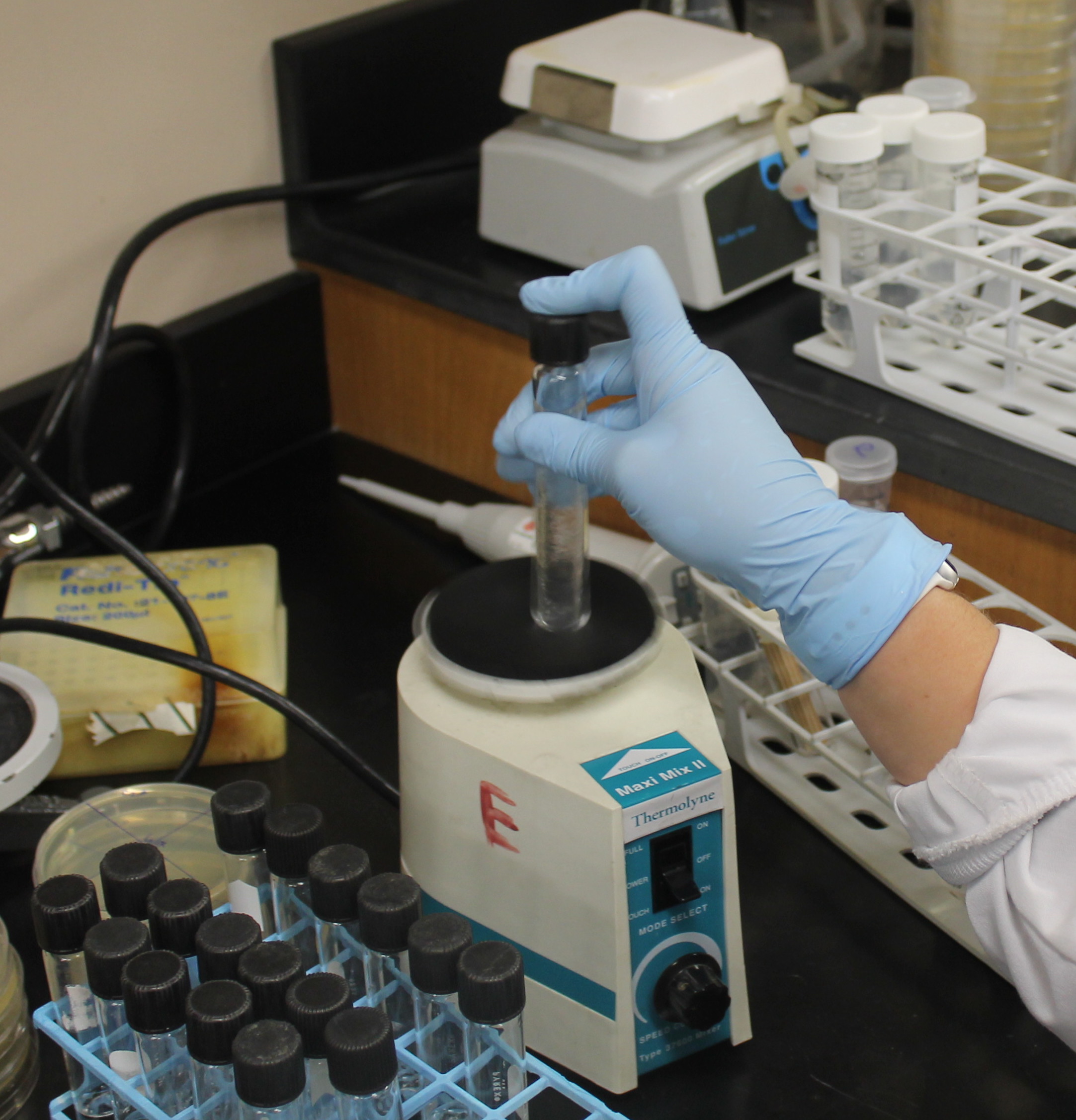} &
\includegraphics[scale=0.04]{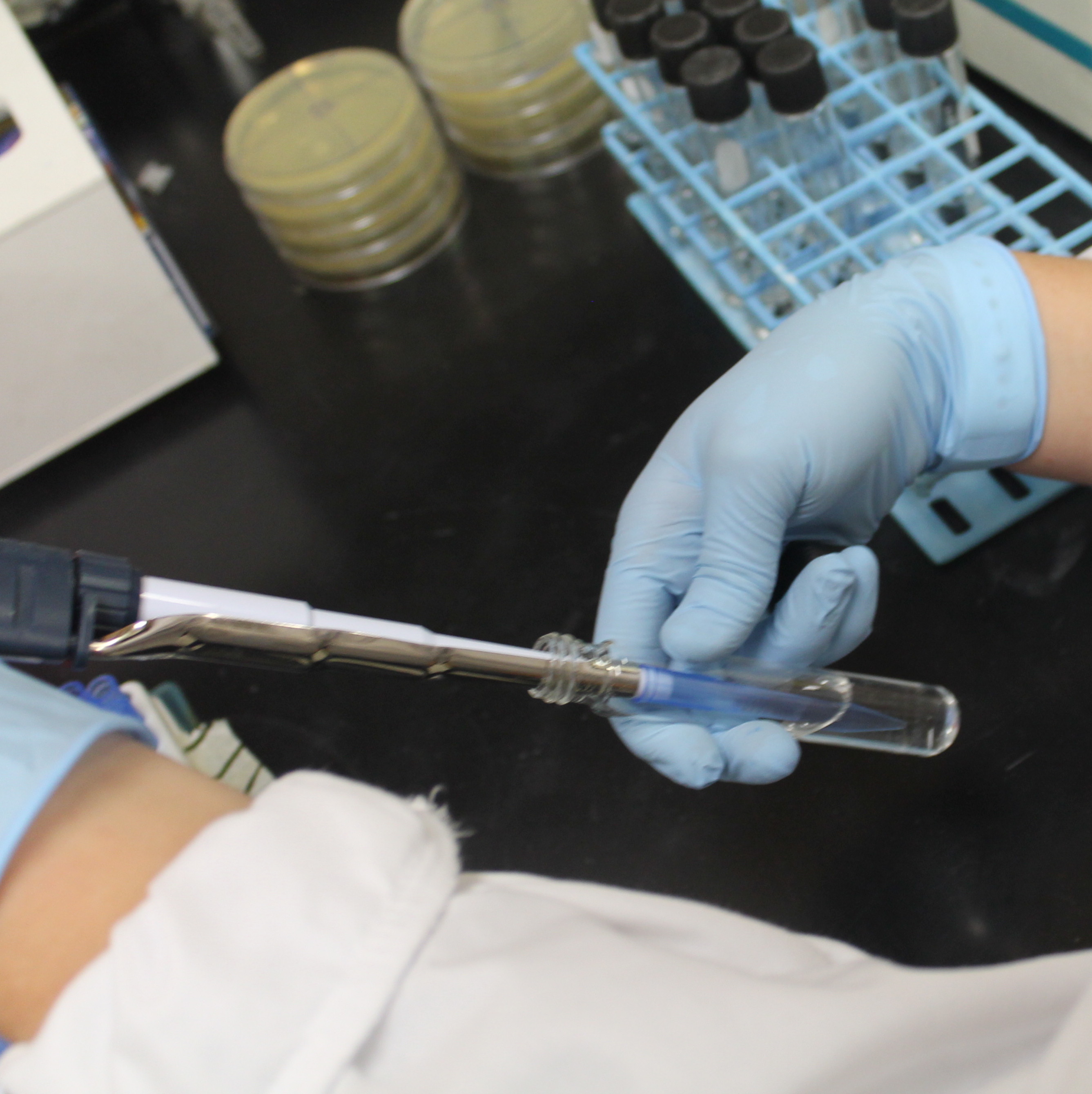} &
\includegraphics[scale=0.043]{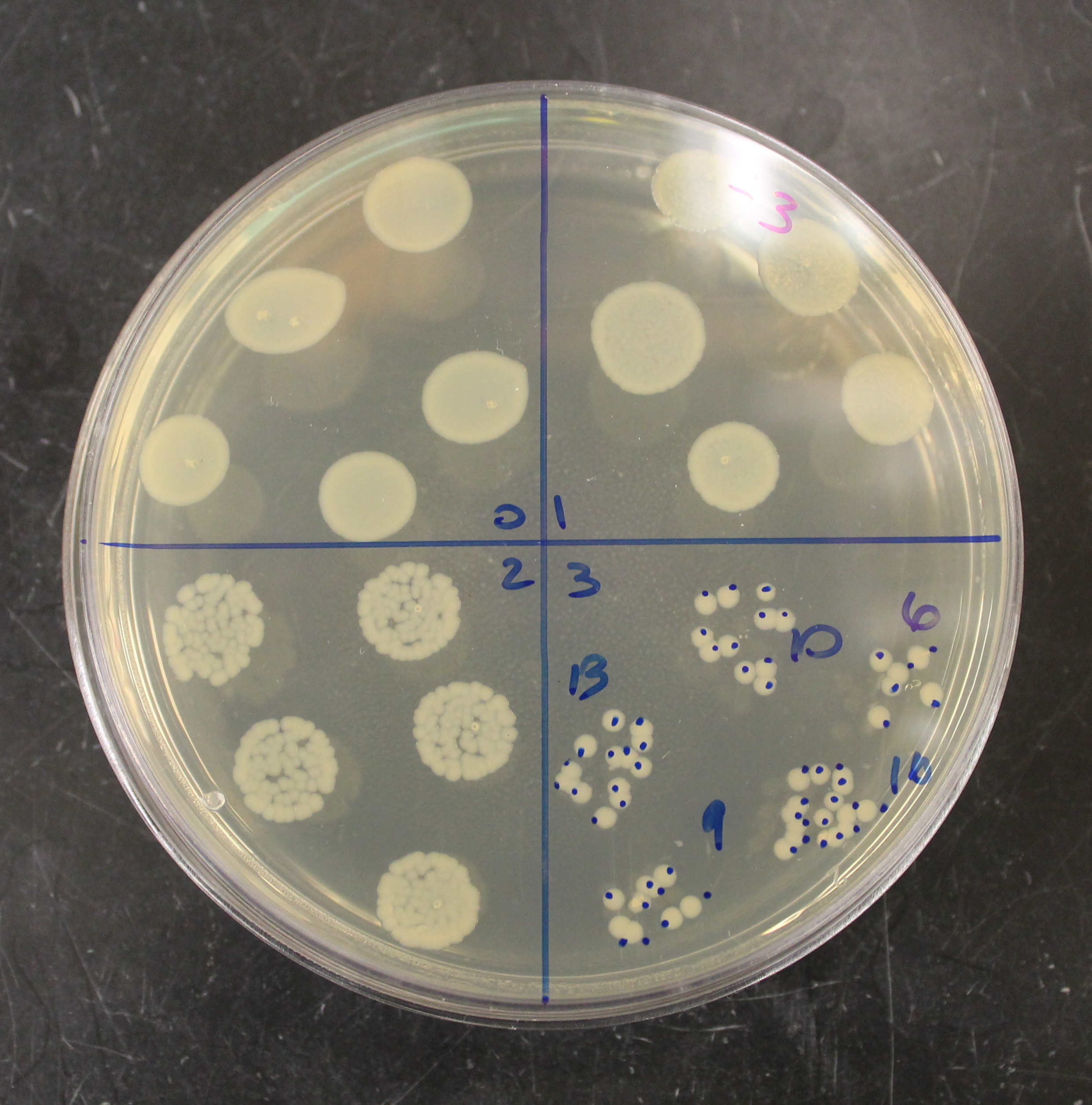}\\
(a) & (b) & (c) \\
\end{tabular}
\caption{\label{fig:CFU_count} (a) Treated samples (here there are $K=3$ repetitions) are transported and placed in dilution
0 tubes (top right) and from these subsequent dilutions are formed (bottom left) by adding buffered water
and then homogenizing with an orbital shaker (center).  (b) From each dilution a volume is taken
and drops are plated using an electronic micropipette. (c) Example of  drops plated in Petri dish
from dilutions 0, 1, 2 and 3.  Dilution 3 (ie. $j=3$) is selected for CFU counting, others are ``too numerous to count'' (TNTC).
Counts for dilution 3 are 13, 10, 6, 9 and 16.  As in the example in Section~\ref{sec:intralab}, $\alpha_0 = 1, \alpha=10$ and
$\alpha_p=10^{3}$ leading to a simple mean estimator of $10.8 \times 10^{6}$.  We use a binomial model to approach this estimation problem formally, and use a Bayesian approach to quantify the uncertainty in estimates of CFU counts.}
\end{center}
\end{figure}

Dilution experiments are an important tool to detect the presence of microbes, even in very low
concentrations, relying on basic microbiology techniques and relatively simple laboratory equipment.  The microbes are first sampled from a natural environment, such as soil or drinking water, or from an engineered bench-top reactor system where they can be grown planktonically or in a biofilm before potentially being exposed to some treatment.
Whatever the case, the sample containing microbes is transferred into a volume $V_0$ of liquid in tube 0, most commonly buffered water \citep{StandardsWater} but any appropriate liquid medium may be used.  It is then of interest to estimate the number of microbes in this tube 0, i.e., the {\it abundance} of microbes in the original sample.  Analyzing the original sample directly may be possible using specialized equipment and more complex lab processes
\citep[e.g. a confocal scanning laser microscope][]{PittsStewart2008, ParkerJASA}.  Instead, subvolumes from tube 0 may be  ``spread-plated'' or ``drop-plated" \citep{Herigstad} onto growth media in a Petri dish to let bacteria grow until distinguinshable by the naked eye (see Figure \ref{fig:CFU_count}(c)).  When plating directly from samples that contain a high density of microbes, it is not possible to identify and count individual colonies, in which case it is necessary to dilute the original volume until only a few colonies can be counted after plating in a Petri dish.   From these counts at some dilution(s), the microbial abundance is inferred in tube 0 and hence in the original sample.  The process is called a {\it dilution series} and is described as follows.

To begin the dilution series, from the volume $V_0$ in tube 0 a subvolume $V$ is taken which is 
diluted at a factor $\alpha$, typically $\alpha = 10$, to form a new volume $\alpha V$ in a new tube
at dilution $j = 1$.  This process is repeated for $j=1,...,J-1$: a subvolume $V$ is taken from the
volume $\alpha V$ in tube $j$, and then diluted by a factor $\alpha$ to form a new volume $\alpha V$
in a new tube at dilution $j+1$.
A smaller volume $u$ is then removed from each tube and then plated onto
growth media in a Petri dish, and kept at optimal environment conditions (e.g., temperature and pH)
to allow any viable microbes to grow and reproduce. Individual microbes or clumps of microbes will start forming colonies in the Petri dish, until,  at a high enough dilution,  they are distinguishable to the naked eye as dots called Colony Forming Units (CFUs).

Estimating microbial abundances from colonies dates at least as far back as the seminal work of Robert Koch in the 1880's
\citep[][p. 9]{Prescott1996}.  These estimates are expressed as CFUs rather than as the \textit{number} of microbes because of  a number of known limitations, the most obvious being the questionable assumption that each colony arises from an individual cell
(\citet[][p.119]{Prescott1996}; see also \citet{Cundell2015}  for a more recent discussion). 
Still, especially for a single microbial species isolated from a consortia in an environmental sample, or for a mono-culture grown in the laboratory, the CFU remains a useful quantitative measure for estimating microbial abundances.  Dilution series for CFU counting are performed routinely in many government, academic and private laboratories for experimentation as well as for testing and public standard compliance \citep[see for example][or a ``dilution series'' search in \url{fda.gov}]{BenDavid2014, FDA2018}.    Indeed, CFU counting is a required international metric for assessing the efficacy of antimicrobial treatments in North America and Europe \citep{ParkerRepro}. 

In a dilution series, for some lower dilutions, too many CFUs may cluster and will be impossible to count and are reported as
``too numerous to count" (TNTC).
For higher dilutions there will eventually be no CFUs (no microbial activity).  Commonly, one dilution is then selected,
namely dilution $j$, $0 \leq j \leq J-1$, for CFU counting,
having a minimum number of distinguishable CFUs per plate or drop, and referred to as the {\it lowest countable dilution}. From this, the microbial abundance in the original sample is to be estimated.
The crudest estimate is the (number of CFUs) $\times \alpha_0 \times \alpha^{j} \times \alpha_p $ , where $\alpha_0 = V_0/V$ and 
$\alpha_p=V/u$ is the ratio of the dilution tube volume $V$ over the volume plated, or drop volume, $u$.  The usual formula used by practitioners is (number of 
CFUs) $\times V_0/u \times \alpha^{j}$, which is precisely equal to the latter.  See Figure~\ref{fig:CFU_count} for an illustration of the process, CFU counting and basic data analysis.

Instead of taking solely the lowest countable dilution, a possible variant is to weightedly average the CFUs across multiple dilutions \citep{Hedges,BAM,Niemela,Niemi,Parkhurst} that is motivated by the Horvitz-Thompson estimator, popular in field ecology \citep{HorvitzThompson}.
\cite{Hamilton2010} argue that the added information is
minimal for common dilution experimental designs.  Our investigation (presented here) leads to this same conclusion, which supports the microbiologist's conventional practice of using data from only the first countable dilution to estimate the microbial abundance in the original sample.

In most situations the abundance of microbes in the original sample spans several orders of magnitude and therefore it is common to estimate a {\it log abundance} that is the $\logten$-transform of the CFU count in dilution 0.  The estimated mean log abundance from statistical analyses can easily be normalized to a mean {\it log density} per unit of the volume or surface area of the original specimen, $S_c$, by simply subtracting by $\logten (S_c)$.   

The efficacy of an antimicrobial treatment is usually quantified by a {\it log reduction} (LR) which is the estimated mean log abundance of microbes that survived the treatment subtracted from the estimated mean log abundance of microbes in a concurrent control.  The microbes in the control samples are subjected to the same conditions as the treated samples with the exception that the controls are subjected to a placebo treatment.  Perhaps not surprisingly, the log abundances of the microbes in the control samples are typically much less variable than the microbes subjected to an antimicrobial \citep{ParkerRepro}.  Hence the constant variance assumption of many statistical models is often violated when analyzing a data set that includes both control and treated samples.

There are two common approaches to analyzing CFU data.  One uses a Poisson likelihood model of the counts \citep{QMRA}, while the other uses a log-normal likelihood model of the counts \citep{Hamilton2013}.  Both maximize the likelihood to provide microbial abundance estimates and quantify uncertainty.  Both also can be extended to handle random effects \citep{Zuur} due to samples being repeatedly collected from the same site, experiment and/or the same laboratory.   Software is readily available to fit either of these types of models \citep[see, e.g.,][]{lme4}.

\citet[][p. 213]{QMRA} argue that a Poisson maximum likelihood estimator (MLE) approach is preferred over the log-normal MLE.  One reason is that the Poisson likelihood naturally deals with zero CFU counts. The obvious downside to the Poisson model  is the requirement that the variance is equal to the mean.  The control data that we present here clearly do not adhere to this restriction.  The Generalized Poisson or Negative binomial are both extensions of the Poisson that allow for the variance to be independently estimated from the mean \citep{Joe2005}. Nonetheless, neither of these Poisson approaches allow one to directly analyze LRs.    

The log-normal MLE approach overcomes the restriction that the mean equals the variance by including separate parameters for the mean and variance. To deal with the differing variability of microbes treated by antimicrobials versus controls, one can aggregate the log abundance estimates into LRs and then apply a normal model to the LRs \citep{Hamilton2013}.
Unfortunately, this approach does not allow one to separately model or estimate the variance among counts on different plates at different dilutions.     When presented with CFU data including a zero or TNTC, a common tactic when using a log-normal likelihood is to substitute in a small value (CFU = 0.5 or 1) for zero and to substitute in the largest count for TNTC (30 for the drop plate method, 300 for the spread plate method).   Many published simulation studies show that as long as there are not too many censored data ($\dot\le 15\%$ of the total data set), the log-normal model has little bias and mean squared error when estimating the mean log abundance of organisms \citep[see, eg.,][]{Clarke,HaasLOD,Singh,EPA_LOD}.

The approach that we present here has several advantages over the conventional approaches just described.

\begin{enumerate}

\item The design and assumptions of the dilution experiments lead directly to our binomial model in~(\ref{eqn:model3}) and~(\ref{eqn:model2}).

\item Any censored data are directly modeled in the binomial likelihood (i.e., no substitution rules are employed)
resulting in zeros and TNTCs handled systematically by the same model.

\item Counts from multiple dilutions are directly incorporated into the model (ie. they are not aggregated together before statistical analysis), which allows us to separate out the variance among counts on different plates and on different dilutions from other sources of variance that contribute to the variance of log abundances and LRs.

\item Our model accounts for clustering of the cells (with the miscount probability $q$)
that violates the assumption that one microbe generates one CFU.  This aspect of the model can also deal with miscounts by the technician who actually counts the CFUs.

\item Instead of summarizing the results from each experiment by a LR and then analyzing the LRs using a normal model, we provide an over-arching hierarchy for the analysis of LRs that has the explicit information regarding the CFUs and dilutions that led to the LR.

\end{enumerate}

The paper is organized as follows.  In section~\ref{sec:model} we build our binomial model from the description of the experimental design.  Using a basic theorem a simplification is obtained leading a straightforward likelihood.  We also describe our Bayesian inference process and a hierarchical modeling strategy to analyze intra and inter-lab data and the MCMC method used.  Two examples are presented in section~\ref{sec:examples} considering real data and in section~\ref{sec:disc} we present a discussion of the paper.

\section{Model and Inference}\label{sec:model}

As explained above, microbial samples are treated with, for example, a chemical agent at some concentration, or water temperature for a specified contact time, and the effect of this treatment  is observed as the  number of CFUs from surviving microbes  using a dilution series.  Commonly, several experiments may be conducted with the same treatment, and within each experiment, multiple repetitions or samples  are also considered.
To ease analysis and notation, we concentrate on one experiment for a single treatment.
We make a comment on the analysis of multiple treatments in section~\ref{sec:disc}.

Throughout, we will refer to the plate or drop from which CFUs were counted only
as `drop'; our model may consider in principle any such volume from which CFUs are counted simply by using a different division factor $\alpha_p$ $=V/u$.

Let $K$ be the number of repetitions of microbial samples to be analyzed in one experiment. Each sample, grown in an original surface area, volume etc. $S_c$, is transported into a volume $V_0$ (typically buffered water) and carefully homogenized in tube 0 of each repetition
$k=1,2, \ldots , K$.  Some fraction $\alpha^{-1} \alpha_0^{-1} V_0$
of the volume $V_0$ in this first tube is taken to tube 1, diluted in
$(1- \alpha^{-1}) \alpha_0^{-1} V_0$ volume 
and carefully homogenized.  This process is repeated to produce tubes
$j= 1, \ldots , J-1$ for each repetition.  Tubes 1 through $J-1$ will have $V = \alpha_0^{-1} V_0$ volume and $\alpha_0 = V_0/V$ is the proportion of volume in tube 0 vs the common volume $V$ in the rest of the dilution series tubes.  In our examples $V_0 = V = 10 ~\text{ml}, \alpha_0 = 1$ (intra-lab example studying heat treatments) and $V_0 = 40 ~\text{ml}, V = 10 ~\text{ml}$ and $\alpha_0 = 4$ (inter-lab example studying bleach treatments).

$D$ `drops' of volume $u$ are then plated from each dilution tube.
For example, for the drop plate method, a total of $D=10$ drops, each with volume $u=10 ~\mu$l, are usually plated.   For the spread plate method, $D=2$ drops, each with volume $u=0.1$ ml or $u=1$ ml, is common.  One or several Petri dishes may be used to grow CFUs from these drops, but we will not distinguish between different dishes from the same dilution tube.
For $j= 1, \ldots , J-1$ the proportion plated is $\alpha_p = V / u$ and for $j=0, \alpha_p \alpha_0 = V_0 /u $.

Let $N^k_j$ be the r.v. representing the number of CFUs in dilution $j$ for repetition $k$ and let $n^k_j$ be any particular realization for it (we use the standard probabilistic notation of upper case being the r.v. and lower case a particular value for it, eg. $P( Y \leq y | X = x)$, the probability of $Y$ being less or equal to the particular value $y$ conditional on $X = x$) .  Let
$Y_{j,i}^k$ be the CFU count in drop $i = 1,2. \ldots , D$, for dilution $j$ for repetition $k$.
Our approach is to build a hierarchical model following the dilution process and the counting
process just described.  Due to homogenization, we can safely assume that $N^k_j$ and $Y_{j,i}^k$ are binomial random
variables. Namely
\begin{align}
Y_{j,i}^k | N^k_j = n^k_j , q     &~~\sim
Bi( n^k_j , \alpha_p^{-1} \alpha_0^{-\delta_{0,j}}(1-q) ) \label{eqn:model3} \\ 
N^k_j | N^k_{j-1} = n^k_{j-1}   &~~\sim Bi( n^k_{j-1} , \alpha^{-1} \alpha_0^{-\delta_{1,j}}) \label{eqn:model2} 
\end{align}
for $j = 0, 1, \ldots , J-1$, where $\delta_{i,j}$ is the Dirac function.
Here $1-q$ is included as an additional probability that each individual microbe in each drop actually does form a distinguishable colony and adds to the CFU count.

The proposed model above brings up an interesting philosophical point.  This binomial model is simply describing the way that the experiment is conducted, as opposed to a Poisson or log-normal.
As explained in the introduction, the above model is in fact describing the experimental design with no further assumptions in the statistical modeling than those already assumed by the experimenters conducting the dilution experiments. These assumptions are that: through careful homegenization, a drop from dilution $1$ has a proportion $\alpha_p^{-1}\alpha^{-1}$ of CFUs from dilution 0 and dilution $j$ has a proportion $\alpha^{-1}$ of CFUs from dilution $j-1$.  Moreover, only a small proportion $q$ (maybe less that 5\%) of CFUs fail to make countable colonies.

Moreover, use of the binomial permits direct estimation of the number of CFUs ($N_j^k$'s) with no need for scaling and alleviates the main issues of the other models: over- and under-dispersion in the Poisson case and the use of substitution rules for 0's and TNTCs in the log-normal case.

To model all repetitions in a single experiment, and because it is common to consider log abundances when $N_0^k$ is large,
we take the usual approach of constructing a model linking all log abundances  $\logten ( N_0^k +1)$, $k=1,2, \ldots , K$, as realizations from a population having a common mean $E$ with some dispersion.  This again reflects/models what is being done in the lab: The intention of performing repetitions is to try to estimate the (log) abundance of the experiment itself, and asses its variability by performing $K$ repetitions under repeatitable conditions.
Accordingly $E$ is interpreted as the mean log abundance for a single experiment, which we will infer, and using a Bayesian approach, it will be taken as a r.v. and its posterior distribution estimated.  However, before describing the details of the latter, we first comment on the following two points.

First, as opposed to usual practices, from the onset we add 1 when calculating the log abundance to properly define it when $N_0^k = 0$, that is, when there are only zero CFUs as occurs when no microbes survive an antimicrobial treatment.
Note that for $N_0^k \geq 10$ (as almost always occurs for control samples), $\logten ( N_0^k +1)$ is already nearly the same as
$\logten ( N_0^k)$, so the interpretation of the log abundance defined as $\logten ( N_0^k +1)$ should be straightforward for these large values of $N_0^k$.   When $N^k_0$ is closer to 0, then a log abundance is less useful.  Instead, one can examine directly $N^k_0$ and its posterior for statistical inference. Still,  Taylor's Theorem shows that $\logten ( N_0^k +1) \approx  N^k_0 $ for small $N^k_0$, which suggests defining the log abundance as $\logten ( N_0^k +1)$
even in this case.    Moreover, this definition will permit us not only to be consistent in all cases, but
also to discuss the limit of detection (LOD) when no counts are detected; see section ~\ref{sec:censored}.

Second, it is common to normalize abundances to the volume or surface area of the original specimen, $S_c$, via $N^k_0/S_c$.  We recommend applying this normalization to convert log abundances to log densities by
$\logten  \left(N_0^k + 1\right)-\logten(S_c) = \logten  \left(\frac{N_0^k + 1}{S_c}\right)$.
This brings up an interesting point that is not well appreciated by microbiologists.  Changing the units via $S_c$ clearly changes the mean log density 
but leaves the variance unchanged (due to the multiplicative property of the log transform).   For example, for biofilm samples, when changing the units from CFU/mm$^2$
to CFU/cm$^2$, $S_c$ decreases by a factor of 100, so that the mean log density increases by 2 but the standard error of the sample mean (SEM) and the SD of the individual log densities remain unchanged.   Hence, any frequentist hypothesis test of the population mean of $ \logten \left( N_0^k + 1 \right)$ that depends on a $t$-ratio of the sample mean to the SEM will always become ``statistically significant" for a drastic enough change in units.  This issue is mitigated when considering the log abundance of microbes compared to a concurrent control via a log reduction (LR), as occurs when assessing antimicrobial treatments.  This is because the  LR is unitless.

Returning to the experimental mean log abundance $E$, the first and nearly default modeling approach would be taking
\begin{equation}\label{eqn:gaussian_model}
 \logten \left( N_0^k + 1 \right) = E + \epsilon_k; \epsilon_k \sim N( 0, \sigma)
 \end{equation}
or $ \logten \left( N_0^k + 1 \right) ~~\sim N( E, \sigma)$.
This model might indeed be appropriate for relatively small $\sigma$, to mantain $\logten\left( N_0^k + 1 \right)$ positive, but the Gaussian model is certainly not well suited in general.  To make a more robust modeling approach, first assume that 
\begin{equation}
\label{eqn:model1_mean}
\mathbb{E} \left[ \left.
\logten  \left( N_0^k + 1 \right) \right| ~ E = e \right] = e ;~~ \text{for} ~~ k=1,2, \ldots , K 
\end{equation}
ie. $\mathbb{E} \left[ \logten  \left( N_0^k + 1 \right) \right] = E$.
Second we introduce $A$ as a dispersion parameter, to stipulate the model
\begin{equation}\label{eqn:model1}
\logten \left. \left( N_0^k + 1 \right) \right| E=e, A=a ~~\sim Ga( a, e a^{-1} ) .
\end{equation}
Here we use the parametrization for the Gamma distribution $Ga( a, b)$ where $b$ is the `scale' parameter, and
therefore the expected value above is precisely $e$ as required.  Moreover, its standard deviation is $\frac{e}{\sqrt{a}}$ and its signal-to-noise ratio (ie. mean over standard deviation or the inverse of the coefficient of variation) is $\sqrt{a}$, representing the unitless dispersion in the model which is specially well suited for positive r.v.'s such as $\logten \left( N_0^k + 1 \right)$.  Note also that the above gamma model correctly generalizes the Gaussian model \eqref{eqn:gaussian_model} since for large $A$ (eg. $A>30$) the gamma distribution is already close to a Gaussian and $\logten \left( N_0^k + 1 \right) \dot{\sim} N( E, \frac{E}{\sqrt{A}} )$ and this is precisely the case when we have relatively small variances (low coefficient of variation), making the Gaussian model appropriate.  Then simply, (\ref{eqn:model1}) generalizes the default Gaussian model in (\ref{eqn:gaussian_model}) to positive only values, only using a different, and perhaps better suited, parametrization.

The specification above in fact creates a hierarchical model and using a simple well known
result we may integrate out the $N_j^k; j=0,1, \ldots , J-1$ and the binomial model in (\ref{eqn:model3}) becomes
\begin{equation} \label{eqn:model3b}
Y_{j,i}^k | N^k_0 = n^k_0 , q   ~~\sim Bi( n^k_0 , \alpha^{-j} \alpha_p^{-1} \alpha_0^{-1} (1-q) ) .
\end{equation}
See further details on this and other technical elements of our model in Appendix \ref{sec:app_hierarchical}.

We use a Bayesian approach to make inferences for the parameters of interest by first stating prior distributions and performing an MCMC to sample from the posterior.

We require prior distributions for $E, A, N^1_0, N^2_0, \ldots , N^K_0$.  We take a pragmatic approach in which the
prior for $N^k_0$ is a discrete uniform distribution from 0 to $10^M$, $E \sim U(0,M)$ and
$A \sim Exp(b)$, $b$ is a scale parameter.  $10^M$ is
a maximum physical capacity of CFUs for the surface with surface area $S_C$ or the volume with volume $S_c$, etc. put to treatment.  In engineered reactor systems especially, experimentalists know the maximum microbial abundance in a sample (i.e., they know $M$).  Indeed, they choose which dilutions to plate based on this knowledge.
$A$ is the shape parameter of the Gamma conditional distribution $f_{N^k_0 | E, A} ( n^k_0 | e, a)$ in (\ref{eqn:model1}).
A simple approach is to take the prior for $A$ as exponential,
resulting with most of its mass from 0 to $2b$.  In the drop plate examples below we use $M=10$ and $b=500$.
This prior parameter for A ($b=500$) was calibrated such that each $\logten \left(N_0^k + 1 \right)$
had nearly the same prior as $E$, that is a $U(0,M)$
(ie. a priori $f_{N^k_0 | E, A} ( n^k_0 | e, a)$ is approximately $U(0,M)$ for all $k$; this may be done
calibrating $b$ with simulated samples from (\ref{eqn:model1})).

We use the t-walk \citep{Christen2010} to produce a MCMC algorithm to sample from the resulting posterior distribution.
The t-walk is a self adjusting MCMC algorithm, that requires the log posterior and two initial points.
In the resulting MCMC in all examples we typically generated chains of length 500,000 with an IAT
\citep{Geyer92} of 50 leading to an effective sample size of roughly 10,000.
In our Python implementation the corresponding computations took 50 seconds on a 2.2 GHz processor.  By simulating
initial values for each $N^k_0$ from its `free' posterior (see Appendix \ref{sec:app_hierarchical})
the burn-in resulted very short indeed in most cases.
The initial value for $E$ is taken as the mean of the $\logten (N_0^k +1)$s and the initial value for
$A$ is taken by simulating from its prior distribution.  Overall, the MCMC is very robust, working nearly unsupervised
in all the examples we tested, including all those presented here.

\subsection{Goodness of fit of the binomial model}

Our binomial model simply follows what is actually done in the la\-bo\-ra\-to\-ry and therefore we claim it models dilution series count data correctly.  For example, it accounts for all extreme or censored data cases.   However, there may be unaccounted sources of variability that could question the appropriateness of the binomial model.  To support our claim that our approach is an overall better model for making the correct inferences, we here compare it to an alternative model.  
Obvious models to compare the binomial model with include the Poisson, negative binomial, or generalized Poisson.
The generalized Poisson or negative binomial are both Poisson mixtures \citep{Joe2005},
and would be good candidates for a comparison but it is not clear how to model the dilution series, as we did in (\ref{eqn:model3}) and (\ref{eqn:model2}) with respect to the parameter of interest $N_0^k$ (the abundance in the original sample) to provide a fair comparison with our model.
Certainly alternative definitions can be attempted to use some Poisson mixture as a model for dilution data, but a straightforward  generalization of our binomial approach is a Beta-binomial distribution.

The Beta-binomial distribution is the result of a mixture of a binomial and a Beta distribution, and as such is a generalization of the binomial distribution.  Namely, if $Y | S=s ~\sim Bi( n, s)$ and $S \sim Beta( \alpha, \beta)$ then $Y \sim BetaBinomial( n, \alpha, \beta)$.
That is, this generalizes the binomial distribution by letting the sucess probability to be random, resulting in a variance that is independent of the mean, whereas for the binomial model, its expected value over variance is always $>1$.

It is clear now how to extend our binomial model in (\ref{eqn:model3b}), namely,
$$
Y_{j,i}^k | N^k_0 = n^k_0 , S=s  ~~\sim Bi( n^k_0 , s)  ~~\text{and}~~  S \sim Beta( s^*\lambda, (1-s^*)\lambda),
$$
where $s^* = \alpha^{-j} \alpha_p^{-1} \alpha_0^{-1} (1-q)$.  Here
$E(S) = s^*$ as required, for any $\lambda > 0$.  That is, instead of setting $Y_{j,i}^k | N^k_0 = n^k_0  ~\sim Bi( n^k_0 , s^*)$ as in (\ref{eqn:model3b})
we let $s^*$ become a parameter, the r.v. $S$, with $E(S) = s^*$, and in this way generalize the binomial model to a Beta-binomial.
Here $\lambda$ is an additional hyperparameter for the above Beta (prior) distribution for $S$. 

We use the Bayesian model comparison machinery, calculating the posterior odds of the Beta-binomial vs the binomial model, namely, the Bayes Factors (BF) of the Beta-binomial in favor of the binomial model \citep[see][for example]{KASS1995}.  In either case, the posterior probability for $N^k_0$ is discrete and its normalization constant is calculated with a sum; this normalization constant is calculated for each of the binomial and the Beta-binomial models.
The posterior probability of each model is proportional to its normalization constant and the BF is the ratio of these posterior probabilities or in fact the ratio of the normalization constants of each model.

We still need to fix $\lambda$ to stipulate the Beta prior distribution for $S$ in each case.   
If $s^* \lambda < 1$ the prior mode for $S$ is at zero and would bias the Beta-binomial model towards zero CFU counts in all cases.
To avoid this we need $s^* \lambda \geq 1$, that is $ \lambda \geq (s^*)^{-1}$.  
We have seen that not restricting $s^* \lambda \geq 1$ leads to a BF of practically zero in favor of the Beta-binomial model, besides cases with zero CFU counts, as expected (results not shown).
Since in most cases $(s^*)^{-1}$ is quite large, then setting $\lambda = (s^*)^{-1} + 1$ represents a neutral choice.

\begin{figure}
\begin{center}
\includegraphics[scale=0.6]{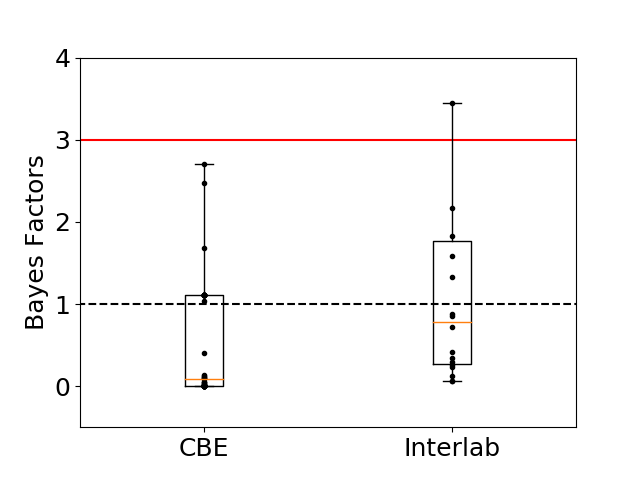}
\caption{\label{fig:BFs}Boxplots of the posterior odds (Bayes Factors, BF) comparing the Beta-binomial vs the binomial model, except for 3 cases with BF $> 4$.
From the 69 data sets analyzed, only 4 had a BF $ > 3$ and 43 had BF $< 1$.  These results fail to indicate clear evidence in favor of the Beta-binomial model but instead seem to favor the binomial model.}
\end{center}
\end{figure}

We performed the Bayesian model comparison on the data sets mentioned in section \ref{sec:examples}. For the 51 CBE and 18 inter-lab individual dilution series count data we found only 4 BF's above 3: one for the former, with a BF = 15 (tube 3 of experiment 70 $^o$C, 10 min), and three for the latter data set, with BF's of 15, 5.5 (from tube 1 of lab 5, tube 1 of lab 6 ) and 3.45.  Leaving out those BF's above 4, the rest of the BF's are plotted in the boxplots in Figure \ref{fig:BFs}.  We are using the common recommendation that a BF above 3 provides positive evidence agains the default model \citep{KASS1995}.  
These results do not provide strong evidence for the Beta-binimial over the binomial model.
Moreover, most of the BF's were below 1 which is evidence in favor of the binomial model over the alternative, more general, Beta-binomial.

\subsection{Censored data, zero counts and level of detection}\label{sec:censored}

As explained in the introduction, the usual practice when counting CFUs is that, besides the actual dilution selected for counting,
the other $Y_{j,i}^k$s are not recorded.  Therefore, the CFUs in the first uncounted dilutions
could be considered as censored data since we know that drops below the selected dilution are
TNTC, and also CFU counts in drops above the selected dilution could also be counted, even if all zero.
However, this added difficulty in recoding and analysis does not seem to add any substantial information
\citep{Hamilton2010}, but this will depend on the particular design, mainly on the dilution parameters
$\alpha$ and $\alpha_p$.  We have confirmed this using our model and the usual drop plate design
using a simulation study (see Appendix~\ref{sec:app_mutidil}); we do not discuss this possibility any further.

\begin{figure}
\begin{center}
\begin{tabular}{c c c}
\includegraphics[scale=0.3]{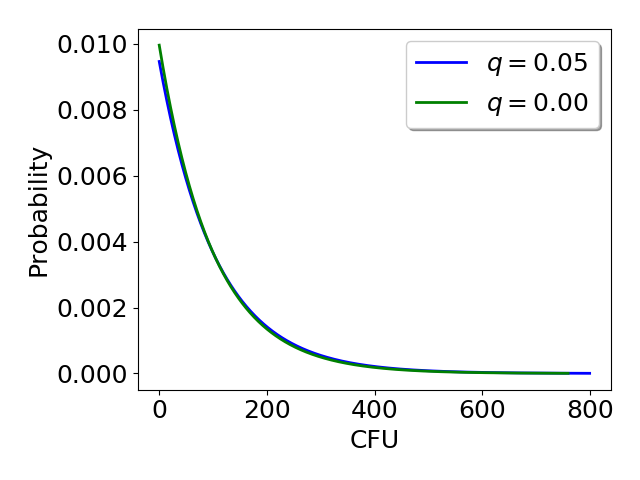} &
\includegraphics[scale=0.3]{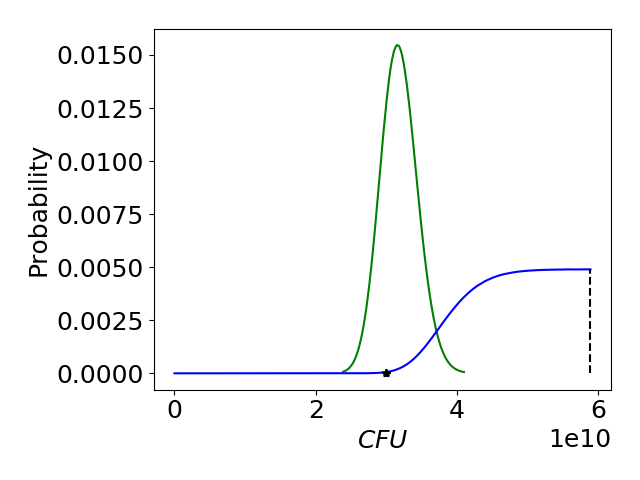} &
\includegraphics[scale=0.3]{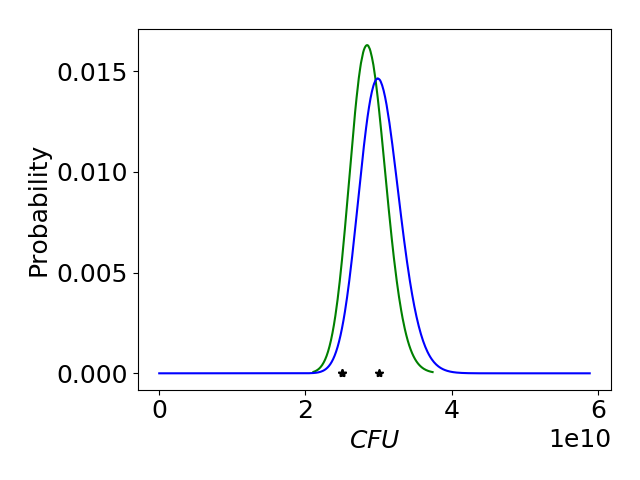} \\
(a) & (b) & (c)\\
\end{tabular}
\caption{\label{fig:Censored}
(a) Posterior distribution for the total CFUs in dilution 0, when zero CFUs
are counted, $\alpha = 10, \alpha_p=1000$ and $D=10$ drops.
A comparison is shown when the miscount probability is $q=0.05$ or $q=0.0$; little difference is observed in this experimental setting.   (b) Example showing all TNTC at the highest dilution, ignoring the censorship setting
$Y_i^k = c$ (green) and properly analyzing including censorship in the likelihood (blue).  Note that in the latter,
the posterior reaches a plateau, since \textit{any} arbitrarily large value is equally likely, and is only truncated
by the prior, ie. a maximal feasible number of CFUs for the coupon surface.  (c) Example of a hybrid case in which
only three out of 10 drop counts where not TNTC, all close to $c=30$, ignoring the censorship (green) and
considering the censorship in the likelihood; already little differences are observed.}
\end{center}
\end{figure}

Accordingly, let $j_k = 0,1, \ldots , J-1$ be the dilution selected for counting (i.e., the lowest countable dilution) for repetition $k$ and to ease notation
let $Y_i^k = Y_{j_k,i}^k$, the actual CFU count in the $i$th drop.  The selected dilution is part of our experimental
design and is taken as the first dilution such that $Y_{j,i}^k \leq c$ ($c=30$ when drop plating, $c=300$ when spread plating).   If zero CFUs appeared even
in the first tube, we let $j_k = 0$ and $Y_i^k = Y_{0,i}^k = 0$.
Dealing with zero counts, that is $Y_{i}^k = 0$ (ie. no CFU counts even at dilution 0), represents no problem and
may be dealt with consistently since (\ref{eqn:model3b}) leads to a likelihood from which a posterior may be
calculated; see Figure~\ref{fig:Censored}(a).

In the case where even at the highest dilution still the CFU count is
above the threshold $c$, that is $Y_{j,i}^J > c$ then the CFU is recorded as TNTC, and we may treat this as right censored data.  The likelihood
in this case is $P[ Y_{j,i}^J > c |	 N^k_0 = n^k_0 , q]$ and this extreme case can also be dealt with, although with a higher computational burden; it involves calculating the binomial cdf in the likelihood.  An example of this saturated data is presented in
Figure~\ref{fig:Censored}(b) and~(c). In the real data examples presented in Sections~\ref{sec:intralab} and~\ref{sec:interlab} no saturated counts are present.

Regarding the miscount parameter $q$, note that for usual drop plate experimental designs $V = 10$ ml, 
$\alpha = 10, \alpha_p=1000$, the success probability in the binomial model is (\ref{eqn:model3b})
$\alpha^{-j_k} \alpha_p^{-1} (1-q)$ which will be quite similar to $\alpha^{-j_k} \alpha_p^{-1}$ for
reasonable miss count probabilities $q \leq 0.1$.
The effect of $q$ will be barely noticeable.  For other experimental settings, the effect of $q$ could
be more important, in which case $q$ could be also considered a (nuisance) parameter to be included in the
posterior, with a tight prior for $q$ small.  However, it would be advisable to make an experimental design
to learn about the miscount parameter $q$, with a reference sample with a known microbial abundance  and
several repetitions.  However, to our knowledge, these hypothetical experiments have not been conducted as yet.
In the examples presented in Section~\ref{sec:examples}
we simply fix $q=0.05$, this should barely have an effect, as seen in Figure~\ref{fig:Censored}(a).
Nonetheless, $q$ is included for overall consistency of our approach.

\begin{figure}
\begin{center}
\includegraphics[scale=0.5]{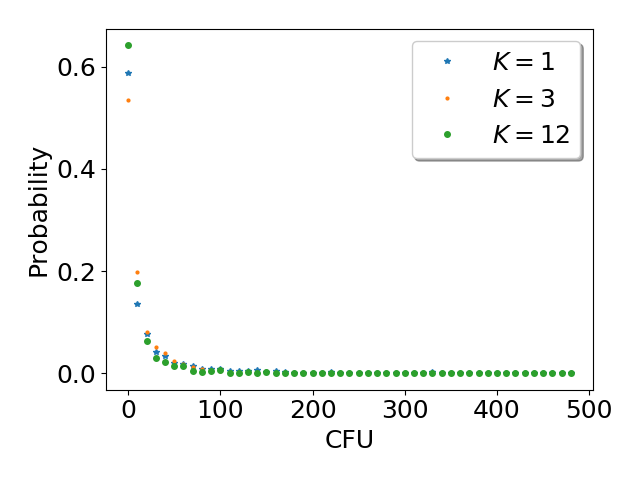}
\caption{\label{fig:LOD} Posterior pmf for $10^E - 1$ (in 10 CFU bins), setting all drop counts to zero,
with $K$ repetitions. Drop plate design $\alpha=10, \alpha_p=1000$, $D=10$ drops.  Lower LODs,
ie. $L_K$ such that $P( 10^E - 1 < L_K |  Y_0^k = 0, k= 1, \ldots K) < 0.95$, are
$L_1 = 110, L_3 = 50$ and  $L_{12} = 30$.}
\end{center}
\end{figure}

Regarding the lower limit of detection \citep[LOD,][]{CurrieOrig},
that is, the minimum number of CFUs that can be detected in
dilution 0 with the chosen experimental design, we may calculate $L_K$ such that
$P( 10^E - 1 < L_K |  Y_0^k = 0, k= 1, \ldots K) < 0.95$, for various repetitions $K$.
That is, setting all drop counts to zero, we define the lower LOD as the 95\% upper quantile $L_K$
of the corresponding posterior distribution for the microbial abundance $10^E - 1$.
In Figure~\ref{fig:LOD} we present examples of these posteriors for the drop plate design
($\alpha=10, \alpha_p=1000$, $D=10$ drops) with $K=1 , 3$ and $12$  repetitions.  The results are
$L_1 = 110, L_3 = 50$ and  $L_{12} = 30$.  A substantial increase in the lower LOD is obtained when increasing from
1 to 3 repetitions, but the increase in the lower LOD is slower then onwards, coinciding with the
current practices of performing $K=3$ repetitions per experiment.   This same approach can be applied to similarly assess the upper LOD for any experimental design (a useful quantity to estimate when some drop counts are TNTC). 
We further comment on LODs in our approach in section~\ref{sec:disc}.

\subsection{Inter-laboratory hierarchical model analysis}

If the treatment has also been repeatedly studied in $L$ different laboratories, each lab will have an independent hierarchical $E$ variable, denoted by $E_l; l=1,2, \ldots, L$. Analogously we may define a global variable
$\mathcal{E}$ and $\mathcal{A}$ where
\begin{equation}\label{eqn:interlab_hie}
E_l | \mathcal{E} = e, \mathcal{A} = a ~~\sim Ga( a, e a^{-1} ); l=1,2, \ldots, L 
\end{equation}
and again this generalizes the default Gaussian approach to positive values.  This also describes what practitioners initially intend to do:  infer an overall $\logten \left( CFU +1 \right)$ across many labs and asses its variability. 
This constitutes an additional hierarchy, where now $\mathcal{E}$ represents the global mean for $\logten \left( CFU +1 \right)$,
for the experiment across laboratories.
The t-walk can be used to generate a chain of posterior samples, as was the case for the hierarchical models within each lab.
We use this approach in the inter-laboratory examples in Section \ref{sec:interlab}.

\section{Examples}\label{sec:examples}

All Python code and data from examples in sections~\ref{sec:intralab} and~\ref{sec:interlab} are available in the supplemental material.

\subsection{Intra-lab analysis}\label{sec:intralab}

Data are taken from a series of experiments performed in the Center for Biofilm Engineering, Montana State University, MT, USA.
Biofilms of {\it Sphingomonas parapaucimobilis} were grown on a cylinder with surface area $S_c=$ 4.52 cm$^2$ and then subjected to different temperatures for varying contact times.  Each temperature and time
combination represents a treatment and each treatment was applied to $K=3$ replicate biofilm samples, as described in \cite{Wahlen2016}.
By sonication the biofilm is harvested from the cylinders into $V_0=10$ ml of buffered water to form dilution $j=0$.  To begin a $\alpha=10$-fold dilution series, next $\alpha^{-1}\alpha_0^{-1}V_0=1$ ml is taken from dilution 0 and then 9 ml of water is added to form dilution $j=1$, etc. up to dilution $j=6$. Ten drops of $u=10 ~\mu\text{l} = 10^{-3}  \cdot 10 ~\text{ml} = \alpha_p^{-1}\cdot V$ each are plated and placed at $36\pm 2 ~^o$C
for $48\pm 2$ h and in turn inspected for CFU formation.   Therefore we have $J=7, \alpha_0 = 1, \alpha=10, \alpha_p=10^{3}$ and $D=10$.

\begin{figure}
\begin{center}
\begin{tabular}{c c}
\includegraphics[scale=0.4]{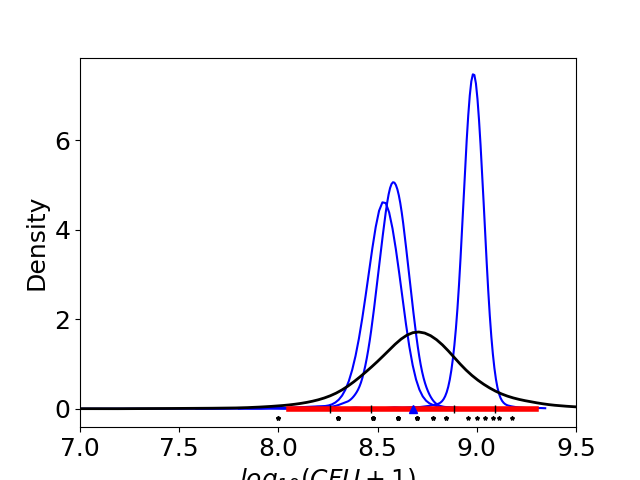} &
\includegraphics[scale=0.4]{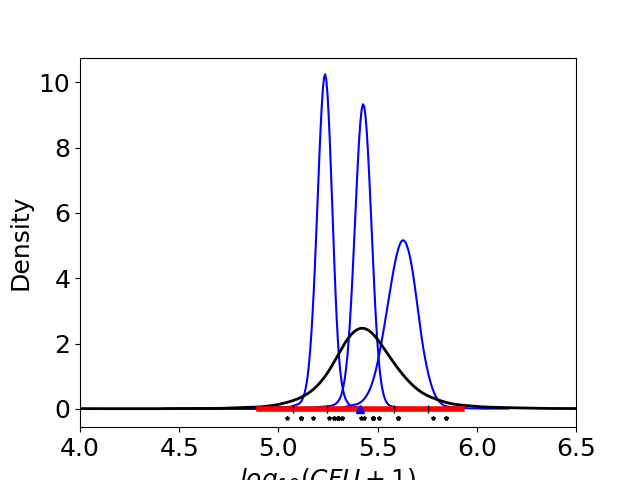} \\
(a) & (b) \\
\includegraphics[scale=0.4]{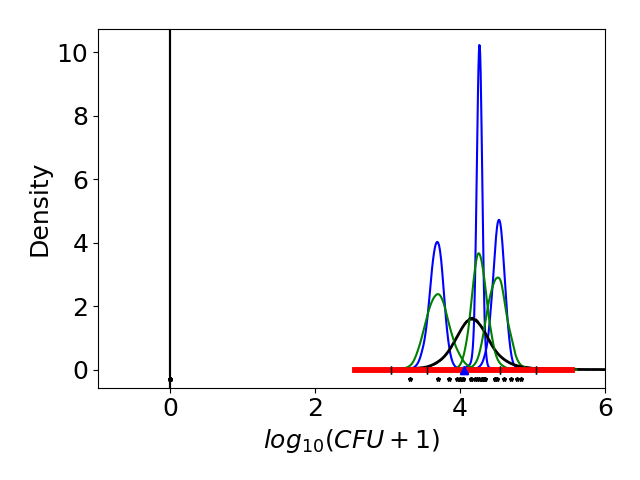} &
\includegraphics[scale=0.4]{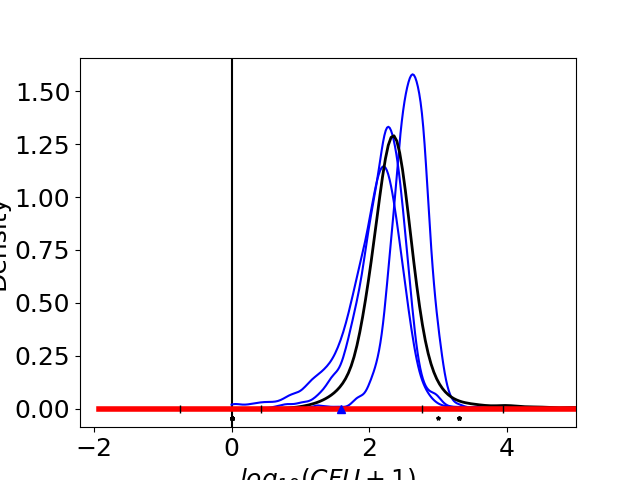}\\
(c) & (d)
\end{tabular}
\caption{\label{fig:PostComp} Posterior marginal distributions for $E$ (thick black) and
$\logten(N_0^k + 1); k=1,2,3$ without hierarchical model (thin blue),
along with the $3\sigma$ intervals and data, for: (a)  treatment room temperature for 15 min the $3\sigma$ interval seems over dispersed. (b) treatment $65 ~^oC$ for 15 min, the classic $3\sigma$
more or less coincides with the posterior variability. (c) treatment $70 ~^oC$ for 10 min, the $3\sigma$ interval looks clearly over dispersed. (d) treatment $75 ~^oC$ for 10 min the $3\sigma$ interval is wrong, covering negative values.}
\end{center}
\end{figure}

In Figure \ref{fig:PostComp} we present the results of our analysis  of 4 treatments:
Room temperature for $15$ min,
65 $^o$C for $15$ min, 70 $^o$C for $10$ min and 75 $^o$C for $10$ min.
Also included are results from the classic simple analysis (using the log-normal approach) of calculating confidence intervals from a mean
and a standard deviation on the estimated log abundances.  The resulting intervals, in general, do not coincide with the variability
in the posterior distributions and in some cases may result in confidence intervals that include negative values;
see  Figure \ref{fig:PostComp}(d).

\begin{figure}
\begin{center}
\begin{tabular}{c c c}
\includegraphics[scale=0.3]{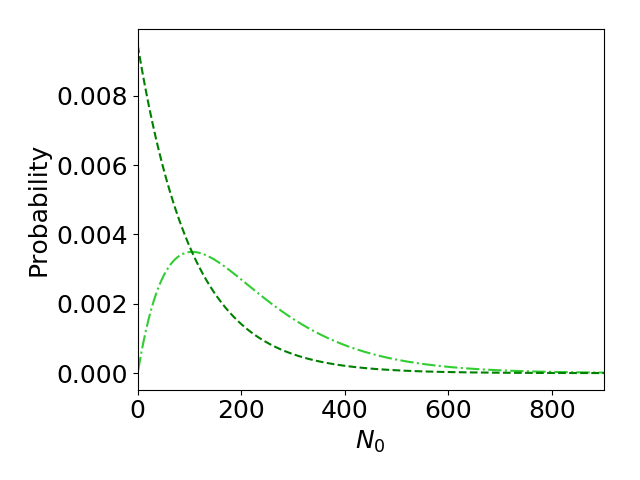} &
\includegraphics[scale=0.3]{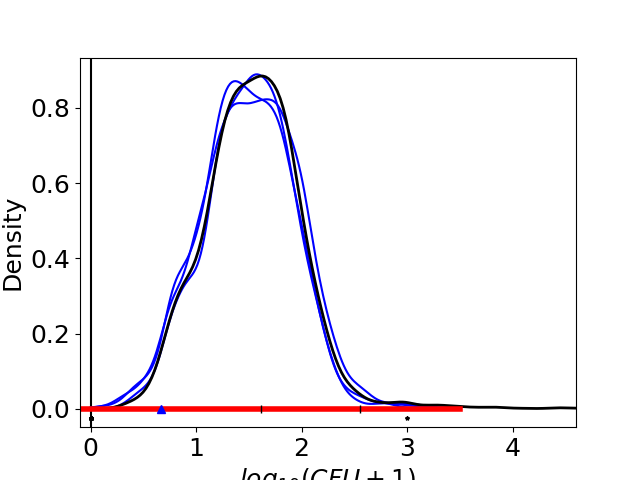}&
\includegraphics[scale=0.3]{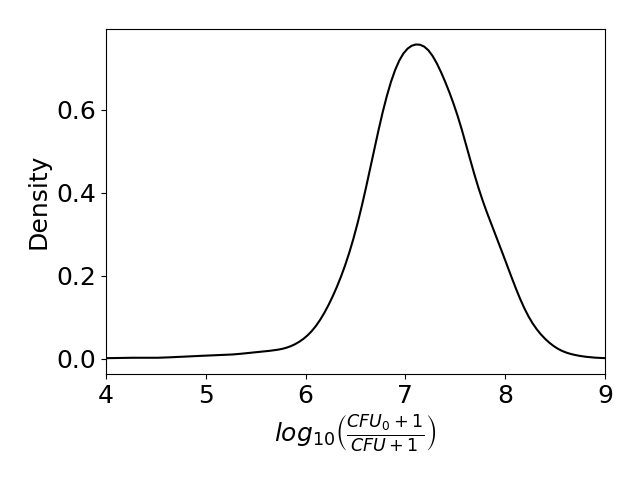}\\
(a) & (b) & (c)\\
\end{tabular}
\caption{\label{fig:PostComp2} Results from treatment $80 ~^oC$ for 2 min:
(a) Individual (`free') posterior pmf's, without hierarchical model, for
$N_0^1$ and $N_0^2$ (dashed, both identical) and $N_0^3$ (dashed and dotted) and (b) these
same posteriors transformed to the $\logten(CFU+1)$ scale as cdf's, the posterior marginal cdf
for $E$ (black) and for $\logten(N_0^k+1)$ (blue solid lines) in the hierarchical model. (c) The log reduction
of the experiment vs. the control (Room Temperature); this is the posterior distribution of $E_0-E$, where
$E_0$ is the $\logten(CFU+1)$ hierarchical parameter for the control experiment, which is in fact
seen in Figure~\ref{fig:PostComp}(a).  Using this posterior distribution we calculate $P(LR > 3 | Data) = 0.9993$.}
\end{center}
\end{figure}

A more extreme example is when most CFU counts are zero.  This case is simply untreatable using a
mean and standard deviation.  For the $80 ~^oC$ and $2$ min treatment, two of the three repetitions had only
zero counts (as mentioned in Section \ref{sec:censored}, we set $j_k=0$ and all $y_i^k = 0, k=1,2$) and the third repetition
had one drop with one CFU only at dilution 0, that is $j_3=0, y_1^3 = 1$ and $y_i^3 = 0, i=2,3, \ldots, D$.
We show the corresponding posterior in Figure \ref{fig:PostComp2}(b).  Moreover, to appreciate the effect of the hierarchical model we include the `free' posterior distributions for the $N_0^k$s (see Appendix \ref{sec:app_hierarchical}), that is not considering the hierarchical model and
each independent posterior for  $N_0^k$ based only on the likelihood based on (\ref{eqn:model3b}); see Figure \ref{fig:PostComp2}(a).
Since for repetitions $k=1,2$ we only had zero counts there is a positive probability that
$N_0^k = 0$, and results in that the mode
of this free posterior is precisely at 0.
However, since repetition $k=3$ had one CFU then
this renders $N_0^k = 0$ logically impossible, and therefore it has zero posterior probability at $N_0^k = 0$,
see \ref{fig:PostComp2}(a).

The corresponding marginal posterior probabilities, now using the hierarchical model, seen in
 Figure \ref{fig:PostComp2}(b), do not match completely with the former posteriors transformed to the
 $\logten( CFU + 1)$ scale.  This is indeed a result of the hierarchical model and the shift in the individual
marginal posteriors is a case of ``borrowing strength'' from one repetition to the other.

In many cases it is desired to study the log abundance of a treatment with respect to its concurrent untreated control, that is the log reduction (LR).
For example, it is common for antimicrobial products to make claims such as ``kills 99.9\%" of bacteria which corresponds to $LR = 3$.
A considerable advantage of using a Bayesian approach is that the LR is analyzed explicitly through its constitutive parts (the controls and treated samples) which can open new possibilities for more comprehensive and goal oriented analyses for dilution series experiments (see also the LR analysis in Section \ref{sec:interlab} and the `activation probability' analysis discussed in Section \ref{sec:disc}). 
Namely, the same hierarchical model may be fitted to the control data obtaining a posterior sample
for the hierarchical log abundance, which we call $E_0$ and inference regarding the LR is simply based
on the posterior distribution of $E_0-E$.
Since we have MCMC posterior samples from both variables, obtaining a posterior sample for $E_0-E$
is immediate; see for an example Figure ~\ref{fig:PostComp2}(c).  Moreover, we may calculate $P(LR > 3 | Data)$ which in this case equals 0.9993 ie. ``with near certainty the $80 ~^oC$ for 2 min treatment killed at least 99.9\% of the biofilm''.

\subsection{Inter-lab analysis}\label{sec:interlab}

\begin{figure}
\begin{center}
\begin{tabular}{c c c}
\includegraphics[scale=0.3]{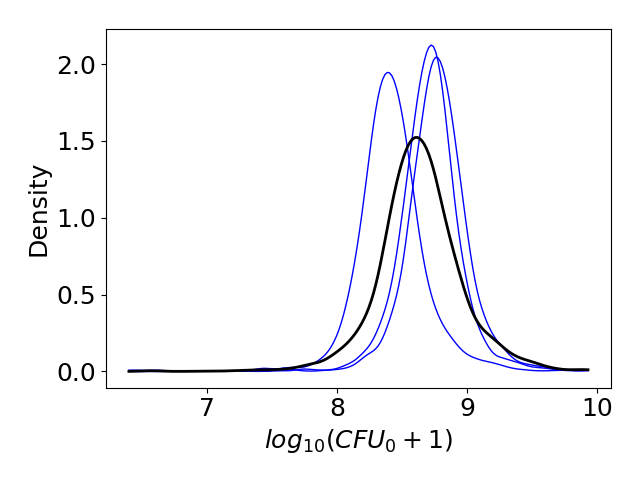} &
\includegraphics[scale=0.3]{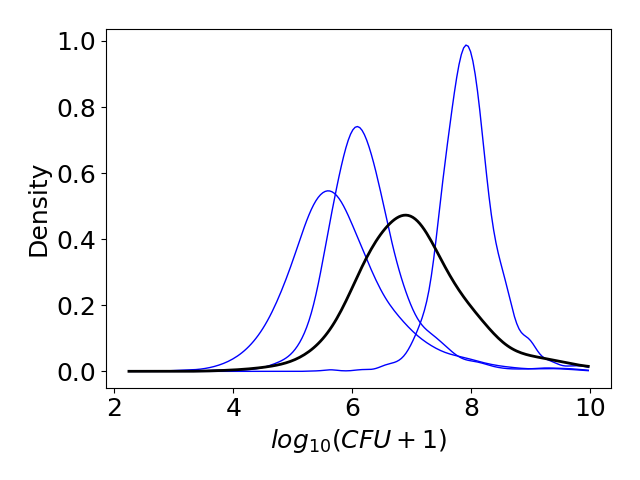} &
\includegraphics[scale=0.3]{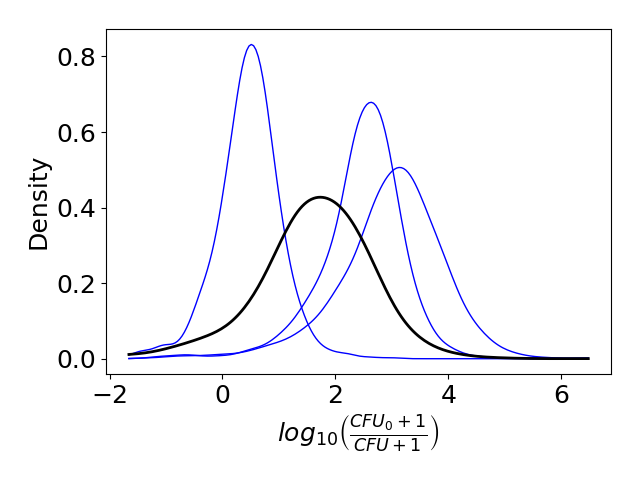} \\
(a) & (b) & (c) \\
\end{tabular}
\caption{\label{fig:InterLab} Inter-lab analysis of results for sodium hypochlorite treatment,
individual $E_l$s for 3 labs not considering
the hierarchical model (blue) and the global $\mathcal{E}$ (black) for the hierarchical model
in~\ref{eqn:interlab_hie}  (a) control, (b) treatment  and (c) LR.}
\end{center}
\end{figure}

In this example we report on results from a multi-lab study of the so-called ``single tube method" \citep{ASTM_STM, ASTM_STM_ILS}, recently standardized by ASTM International (\url{https://www.astm.org/}), to test antimicrobial efficacy against biofilms of {\it Pseuodomonas aeruginosa}.  In this example we focus on 3 labs from the inter-laboratory study of the single tube method, and a single experiment from each lab.  In each experiment, we analyze $K=3$ control samples and $K=3$ samples treated with a low concentration of   bleach (i.e., 123 ppm of sodium hypochlorite) against biofilms.   The biofilms in the labs were harvested into a
$V_0 =40$ ml volume to form dilution 0.  As in the previous Section, to begin a $\alpha=10$-fold dilution series, $\alpha^{-1}\alpha_0^{-1}V_0 =$ 1  ml is taken from dilution 0 
and then  9 ml of water is added to form dilution 1, etc. up to dilution 6 (so there is a volume $V=$ 10 ml in each dilution tube).
Two drops of $u=100 ~\mu$l each are spread plated, and placed at $36\pm 2 ~^o$C
for $26\pm 2$ hr and in turn inspected for CFU formation.   Therefore we have $J=7, \alpha_0 = 4, \alpha=10$, $\alpha_p=10/0.1 = 100$ and $D=2$.  Figure \ref{fig:InterLab} presents the results of our analyses of the log densities ($\log_{10}(CFU +1)$) for each of the control and bleach treatments, and also the LRs.

Unlike other fields of science where there is a demonstrable lack  of reproducibility assessments, in the field of antimicrobial science, many organizations have pooled resources to quantifiably assess the reproducibility, across different laboratories, of methods that assess antimicrobial efficacy \citep{ParkerRepro}.  The reproducibility of test results of antimicrobials is of paramount importance for public health: regulators want to keep ineffective products out of the marketplace, and producers of highly effective products want to bring their products to market.
All stakeholders want laboratory methods that accurately and reproducibly make decisions regarding which products are effective.

The reproducibility issue seems apparent in this example since the results from the LRs for this treatment, as seen
in Figures~\ref{fig:InterLab}(b) and~(c) span $4 - 5$, orders of magnitude.  In a classical setting an analysis of variance (ANOVA) is performed on the dilution series data from multiple labs, but that inherits the problems of the log normal model already mentioned.  In our setting, a formal reproducibility question may be put forward and then assessed with a posterior distribution.  A first choice would be to compare the models for the mean log abundnce $E_l$ for each individual lab versus the full hierarchical model for the global mean log abundance  $\mathcal{E}$, calculating the posterior distribution of each case, for example comparing the individual lab LR's $LR_l = E_l^0 - E_l$ vs. the global LR
$\mathcal{LR}  = \mathcal{E}_0 - \mathcal{E}$; see Table \ref{tab:rep} for a tentative discussion.
This approach is one possibility, but a more in depth analysis is needed to address other reproducibility questions that stakeholders may phrase.   The best approach is to establish the necessary posterior probabilities and, to provide guidance for the  involved decisions to be taken, to maximize posterior expected utilities.  Our Bayesian setting opens up these possibilities, but not without further effort.

\begin{table}
\caption{\label{tab:rep} Further investigating the reproducibility issues in our inter-laboratory example, according to our hierarchical posterior, the  global log reduction $\mathcal{LR}$ is expected to be within 1.7 logs of the individual $LR_l$s for each lab (row 2).  That is, the pooled hierarchical model seem to work well, however, its posterior distribution spans more that 5 orders of magnitud, as seen in Figure~\ref{fig:InterLab}(c).  This lab to lab inconsistency is also suggested in the individual posterior $P(LR_l > 3)$ (row 3), spanning from close to zero to $\frac{1}{2}$, while the inter-lab posterior LR is in fact $P[ \mathcal{LR} > 3 ] = 0.0646$.}
\begin{center}
\begin{tabular}{l | c | c | c}
 Laboratories                        & $l=1$ & $l=2$ & $l=3$ \\  \hline
 $E[ | \mathcal{LR} - LR_l | ]$ & 1.6443 & 1.4053 & 1.2654 \\ \hline
                      $P(LR_l > 3)$ & 0.5348 & 0.0014 & 0.2257 \\ \hline 
\end{tabular}
\end{center}
\end{table}

\section{Discussion}\label{sec:disc}

A key aspect of our model is the use of a binomial likelihood ($Bi(N_0,\cdot)$) with the parameter of interest being $N_0$, the abundance of microbes in the original sample.  We prefer the binomial likelihood because it models what the technicians actually do in the lab.  This is in stark contrast to other common approaches (e.g., the log-normal and Poisson approaches) that provide only an approximation, and markedly different from the case where the statistician imposes an experimental design solely for convenience of the statistical analysis.

This work provides contributions to both the statistical and applied sciences.  To the former, we show how to apply a Bayesian hierarchical model with a binomial likelihood  to estimate and quantify uncertainty about microbial abundances ($N_0$) from dilution series experiments.  The binomial likelihood has a rich history analyzing microbiological data in the case when, instead of CFUs, only a presence/absence response is measured from each tube in a dilution series from which $N_0$ can be estimated using MLE theory \citep{McCrady,Cochran,Garthright}.  Interesting, this MLE approach is called the ``most probable number" technique, coined by McGrady  in 1915 before MLE theory had been developed \citep{HamiltonMPN}.  Our contribution is the first time to our knowledge that CFU data have been modeled with a binomial likelihood.   Regarding our contribution to the applied sciences, we provide a sound alternative to the log-normal and Poisson modeling approaches that are commonly applied in the analysis of CFU data \citep{Hamilton2013,QMRA}.   We have shown that these common approaches have serious deficiencies when modeling CFUs.  For example: in the Poisson case, the restriction on the variance does not hold for CFUs collected from control conditions (that tend to exhibit high abundances and low variability); and in the log-normal case, it is not possible to deal with 0's and TNTC's (without \textit{ad hoc} substitution rules).
The Negative binomial or Generalized Poisson models have  been used to extend the Poisson model \citep{Joe2005}.   Instead, we present a Bayesian approach with a binomial likelihood that allows us to directly estimate the abundance of microbes without a severe constraint on the variance like the Negative binomial or Generalized Poisson would do, and unlike these approaches, our approach can directly incorporate data from multiple dilutions, directly analyze log reductions from the CFUs recovered from  control and treated samples that have different levels of variability, and account for miscounts \citep[recently a shifted Poisson model has been suggested to also handle miscounts][]{BenDavid2014}.   Furthermore, while there is readily available software for application of mixed effects models with either a log-normal or Poisson likelihood to assess reproducibility \cite[see, e.g.,][]{lme4}, we are not aware of similar extensions for the Negative binomial or Generalized Poisson approaches.   We show how our approach may be applied for reproducibility assessments, by comparing the posterior for each individual lab to the posterior for the  population of all labs.   While this is an area of active research, the results we have presented seem promising.

\begin{figure}
\begin{center}
\includegraphics[scale=0.5]{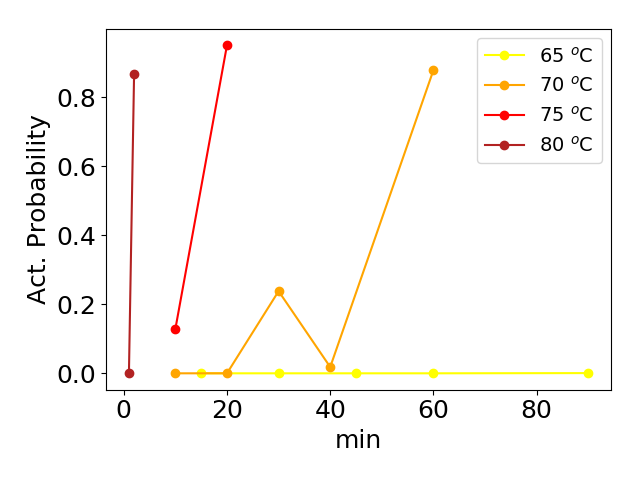}
\caption{\label{fig:activation} ``Activation Probability'' $P( E < e_h \mid \bY = \by )$ with threshold $e_h = 2$, ie. 100 CFUs, for the data described in section~\ref{sec:intralab}.  Coupons were treated at different temperatures and times; at 65 $^o$C the log reduction was not achieved even over a 90 min exposure, whereas at 80 $^o$C the Activation Probability is over 0.8 with an exposure time of only 2 min.  There is a clear difference between 70 $^o$C and 75 $^o$C where similar results are obtained at 20 min at 75$^o$C and only after 60 min for 70 $^o$C.}
\end{center}
\end{figure}

A point not directly addressed in this paper is the analysis and comparison of multiple treatments, since here we concentrated on the novel Bayesian binomial approach using hierarchies for multiple repetitions and multiple laboratories for a single treatment.  The usual objective in analyzing a series of treatments is to fit a regression model to test and/or predict their effect
\citep[see for example][]{Wahlen2016}.
One of the main benefits of performing a Bayesian inference is that we may ask as many questions of interest as we require regarding our parameter of interest, and these are answered with its corresponding posterior probability.
A simple initial approach, that exploits our Bayesian analysis, would be to calculate the actual posterior probability of the desired result, and compare such posterior probabilities across treatments.
For example, commonly in microbiological experiments we are interested in the number of surviving microbes after
treatment. This translates in calculating an `activation probability' of a treatment intended to kill microbial activity, ie. $P( E < e_h \mid \bY = \by )$ for $e_h$ some (small) agreed threshold.  If compared to a control
$E_0$, then we could calculate the posterior probability of a log reduction above some
threshold, that is $P( LR = E_0 - E > l_h \mid \bY = \by, \bY_0 = \by_0 )$ etc., see Figure~\ref{fig:activation} for an example.   This approach lacks the formal predictive approach of fitting a regression model using covariates.

Although not trivial to do, a regression model may indeed be incorporated using a Bayesian approach, embedded in our hierarchical model and binomial likelihood, in a multi-experiment multi-lab scenario.  These interesting possibilities are the focus of future research.  In general, using a Bayesian approach opens the door to a formal Uncertainty Quantification (UQ) approach to the analysis of dilution series data, by exploiting the posterior distribution obtained and the ease in calculation of expectations or posterior probabilities given the MCMC sample obtained.

Instances of 0's and TNTCs relate directly to the lower and upper limits of detection (LOD) for the process used to generate CFUs \citep{CurrieOrig}.  The process includes how the dilution experiment was conducted such as the dilution factors ($\alpha_0$ and $\alpha$), which dilutions were plated ($J$), and the volume plated ($u$). The process also includes properties about the method used to harvest the bacteria from the sample (e.g., sonication, scraping, stomaching or a RODAC plate), the method used to disaggregate/homogenize the bacteria into single cells in the original volume $V_0$ and the environmental conditions used to incubate the bacteria after plating.
In microbiology, it is common to refer to  either $0.5/S_c$ or $1/S_c$ as the LOD of a CFU counting method.  This is unfortunate since neither of these values necessarily are associated with a high level of statistical confidence or probability regarding how many viable microbes survive in the sample.  
In our analyses (Figure~\ref{fig:LOD}) we defined the LOD of the process used to generate CFUs as the lowest microbial abundance that can survive in the sample with probability 0.95 when there were no CFUs recovered from the sample.  From the Bayesian analyses that we present, this calculation is straightforward: the LOD is simply the 95$^{th}$ percentile from the posterior distribution for $N_0|Y=0$;
see Figures~\ref{fig:Censored}(a) and~\ref{fig:LOD}.  Our hierarchical model also easily and consistently can provide the LOD as a function of dilution experiment design (i.e., $J, \alpha_0, \alpha$ and $\alpha_p$) and the number
of repetitions $K$ by using the posterior for $10^E -1$.  For example, in Figure ~\ref{fig:LOD}, it is seen the expected result that the LOD decreases as the number of repetitions increases.  The LOD for a particular experimental design can be calculated, beforehand, at minimal computational cost, to asses if the proposed design meets any LOD requirements.

Regarding the LOD or any other characteristic desired in an experiment, an intensive search can be conducted to optimize the design parameters in Bayesian analyses \citep{Weaver2016, Huan2014, Christen1998}.  This, however, involves calculations of far higher costs, commonly needing parallel computing and other sophisticated software and numerical analysis resources.  Given a design, parameters are simulated from the prior and in turn synthetic data sets from the model, to calculate the corresponding posterior for quantities of interest to asses the expected information gain in data from the design.  This then needs to be repeated from a set of candidate designs to find an optimal design.  We leave this interesting dilution experiment design possibility for future research.

\section*{Acknowledegments}

\acknow

\appendix

\section{Details on the hierarchical model}\label{sec:app_hierarchical}

\begin{figure}
\begin{center}
\begin{tabular}{c}
\includegraphics[scale=0.6]{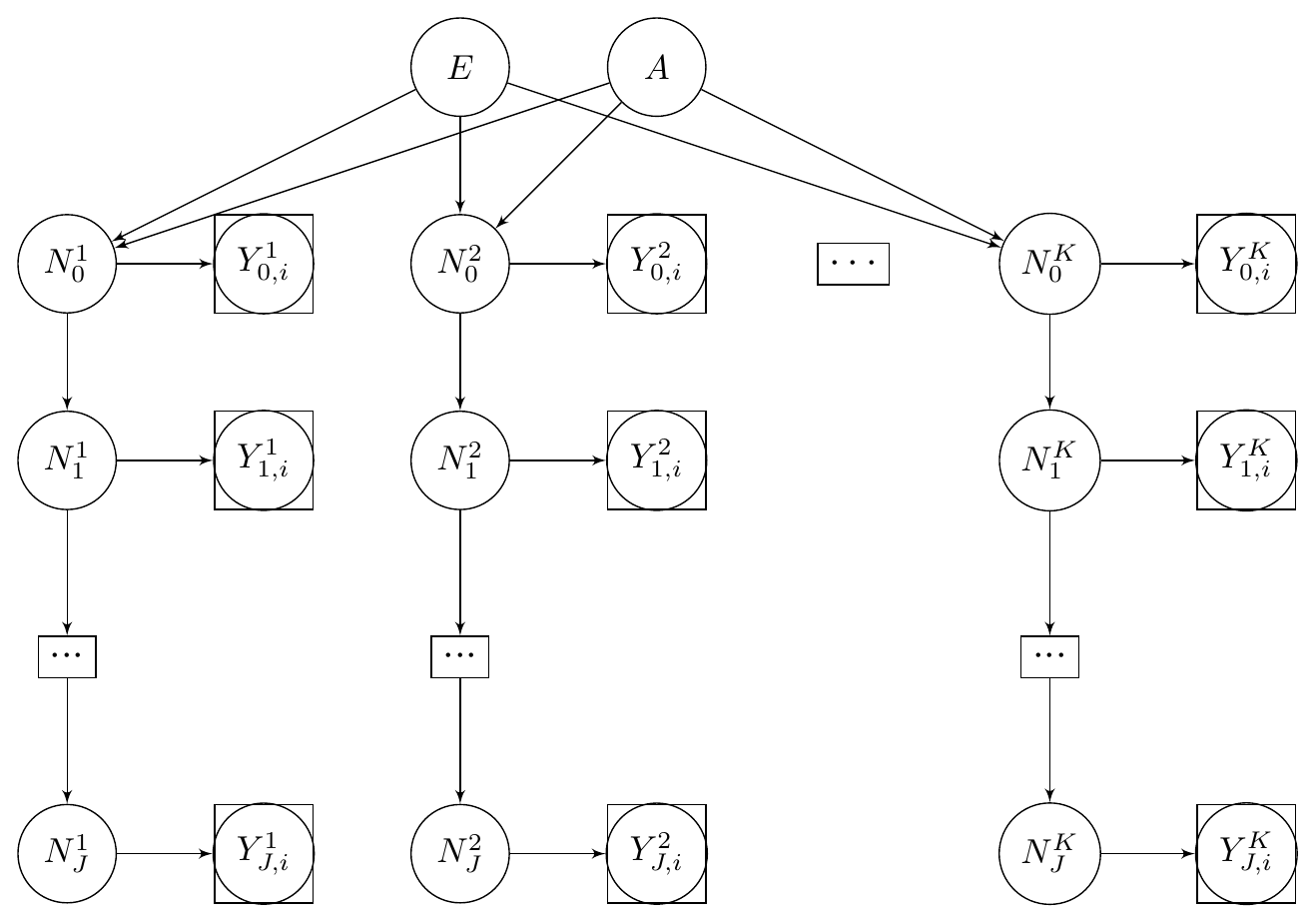} \\
\\
(a) \vspace{0.2cm} \\
\includegraphics[scale=0.6]{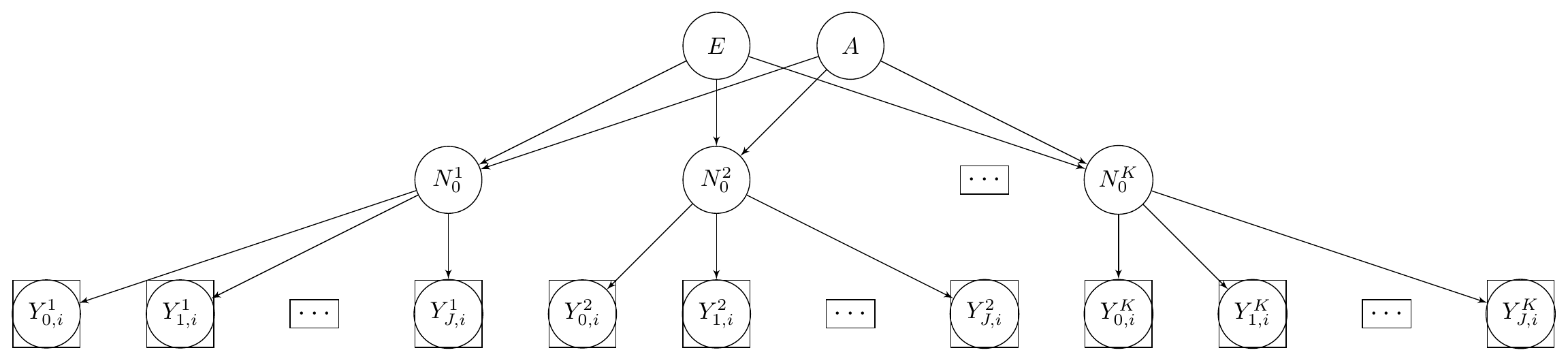} \\
\\
(b)
\end{tabular}
\caption{\label{fig:DAGs} (a) Directed Acyclic Graph (DAG) representing our hierarchical model and
(b) DAG representing our model once the $N^k_j$s have been integrated out, depending now only
on the $N^k_0$s.  Circle nodes are r.v.'s to be estimated, circle and square nodes are r.v.'s that
are observables (the CFU counts).}
\end{center}
\end{figure}

Our hierarchical model may be summarized with the Directed Acyclical Graph (DAG) depicted in Figure~\ref{fig:DAGs}(a).
As mentioned in the main text, using the following well known
result we may integrate out the $N_j^k; j=0,1, \ldots , J-1$.  That is, if $Z \sim Bi( n, p_1)$ and $X | Z=z ~~\sim Bi( z, p_2)$
then $X \sim Bi( n, p_1 p_2)$.  Using it recursively (\ref{eqn:model3}) becomes for $j = 0,1, \ldots, J-1$
\begin{equation} \label{eqn:AppAmodel3b}
Y_{j,i}^k | N^k_0 = n^k_0 , q   ~~\sim Bi( n^k_0 , \alpha^{-j} \alpha_p^{-1} \alpha_0^{-1} (1-q) )
\end{equation}
and the corresponding DAG is as shown in  Figure~\ref{fig:DAGs}(b).  This substantiates the equation \eqref{eqn:model3b} provided earlier.   Since considering $N^k_0 = 0$ is important
and may indeed happen note that it must be the case that $P( Y_{j,i}^k = 0 | N^k_0 = 0 , q) = 1$.  Accordingly
we define $Bi( 0, p)$ as the Dirac's delta at zero.

Note that, strictly speaking,
$\logten \left(N_0^k + 1 \right)$ is a discrete variable and we are modeling it here as a continuos r.v.
The binomial model may be changed to accept a non-integer ``number of trials'' $N_0^k$
using gamma functions in the combination function, having as a particular case the conventional
binomial pmf
(this we do in our implementation providing no further details here).  Then for mathematical convenience
$\logten \left(N_0^k + 1 \right)$ is modeled as a continuos r.v. while still $N_0^k$ is taken discrete.

Moreover, by ignoring the hierarchy involving $E$ and $A$  we may calculate the posterior
distribution of each $N^k_0$ independently.  In this case, since it is a single discrete parameter,
calculating its posterior pmf is straightforward, constructing a likelihood from (\ref{AppAeqn:model3b}).
For illustration purposes and comparisons this is done
in Section~\ref{sec:censored} and in Figure~\ref{fig:PostComp2}(a).  We call this the \textit{free posterior} for
$N_0^k$, the microbial abundance in the $k^{th}$.

The full set of parameters are $E, A, N^1_0, N^2_0, \ldots , N^K_0$.  Only binomial and Gamma densities
are involved and therefore the corresponding likelihood is simple to write and calculate.
The likelihood, indeed since repetitions are exchangeable, is
\begin{equation}\label{eqn:likelihood}
f_{ \bY  | E, A, N^1_0, N^2_0, \ldots , N^K_0 } ( \by | e, a, n^1_0, n^2_0, \ldots , n^K_0 ) =
\prod_{k=1}^K \left\{ \prod_{i=1}^D f_{Y_i^k | N^k_0} ( y_i^k | n^k_0 ) \right\} f_{N^k_0 | E, A} ( n^k_0 | e, a)
\end{equation}
where $f_{Y_i^k | N^k_0} ( y_i^k | n^k_0 )$ is the corresponding binomial pmf stated in
(\ref{eqn:AppAmodel3b}) and $f_{N^k_0 | E, A} ( n^k_0 | e, a)$ arises from (\ref{eqn:model1}),
$\bY$ and $\by$ are the r.v.'s of observed CFU counts arranged in $K\times D$
matrices such that $\bY = (Y_i^k)$ and $\by = (y_i^k)$.  $f_{N^k_0 | E, A} ( n^k_0 | e, a)$ is established
using the fact that if $\logten(Z) \sim Ga( a, b)$ then
$f_Z(z) = \frac{b^{-a}}{\Gamma(a)} (\logten(z))^{a -1} y^{-(b \log(10))^{-1} - 1}$.

\section{Multidilution data analysis}\label{sec:app_mutidil}

\begin{figure}[h]
\begin{center}
\begin{tabular}{c c}
\includegraphics[scale=0.45]{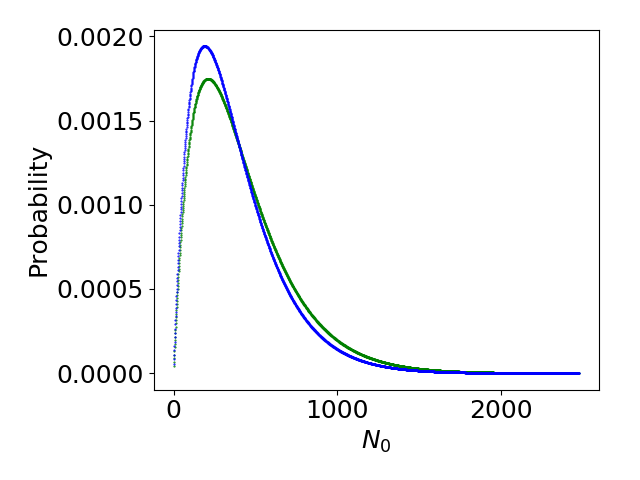} &
\includegraphics[scale=0.45]{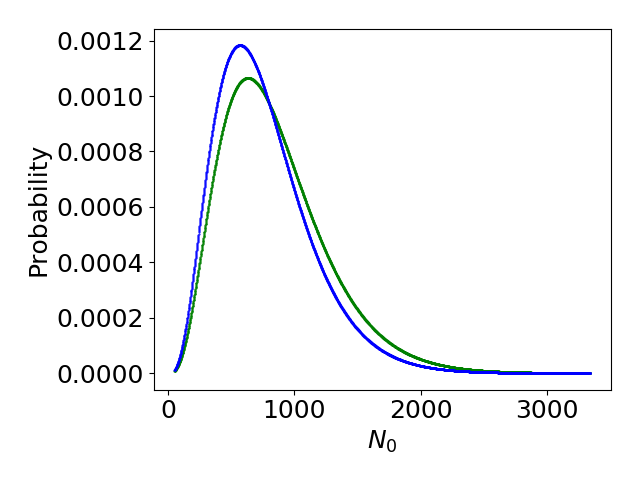} \\
(a) & (b) \\
\end{tabular}
\caption{\label{fig:AllCounts} Expected posterior distribution when all dilution counts are considered (blue)
and when only the first dilution counts are considered (green), for (a) drop plate design,
with true value $N_0=500$, and (b) spread plate design, with true value $N_0 = 50$.}
\end{center}
\end{figure}

We present results on whether it is worth analyzing CFU counts over all dilutions vs over only the first
countable dilution.  This question is difficult to answer in complete generality.
We will then concentrate on a two common designs that correspond to data in Sections~\ref{sec:intralab}
and~\ref{sec:interlab}.  Namely, $\alpha_0 = 1, \alpha = 10, \alpha_p=1000, J=6, c=30$ for the drop plate data in
Section~\ref{sec:intralab} and $\alpha_0 = 4, \alpha = 10, \alpha_p=100, J=6, c=300$ for the spread plate data in
Section~\ref{sec:interlab}.

We study the extreme case when the first dilution can be counted, that is all counts are below the TNTC threshold $c$,
and there is only one repetition ($K=1$).
We fix a true value for $N_0^1$ and simulate data at all dilutions using the binomial model in (\ref{eqn:AppAmodel3b}).  Then we calculate
the discrete posterior pmf of $N_0^1$ and repeat the process averaging over all resulting posteriors over 120
simulated data sets.
The process is done calculating the posterior when only the first dilution is used and when all dilution
counts are considered in the calculation of the posterior.  For the true value of the abundance $N_0^1$ we considered
only a set of representative values for both designs, below a maximum to maintain expected simulated counts below $c$.  Namely,
$N_0^1 =$ 500, 5,000, 10,000, 20,000, 30,000 for the drop plate data design and
$N_0^1 =$ 50, 500, 1,000, 2,000, 3,000 for the spread plate data design.
The posteriors with the most noticeable differences, although still quite low, were at
$N_0^1 = 500$ and $N_0^1 = 50$, respectively; we present these posteriors in Figure~\ref{fig:AllCounts}.
In all other cases the differences in posteriors were even smaller (not shown), suggesting that the added difficulty in counting, recording and analyzing CFUs at all dilutions
is not worth the expected information gain, and therefore we recommend only to record and analyze the first countable dilution.

The reason that the posteriors are so similar can be
seen in Figure~\ref{fig:Censored}.  The likelihood for TNTCs in dilution $j$
becomes flat and provides no information for higher values than $2 c ~ \alpha^j \alpha_p$.
If there are countable drops for dilution $j+1$, the likelihood  for these data is placed at
$\alpha$ fold higher values well beyond $2 c ~ \alpha^j \alpha_p$ and therefore including the
TNTCs likelihood in dilution $j$ will not have any significant effect on the posterior,
at least for the common case where $\alpha = 10$.
Then including the censored likelihood terms for TNTC's will not add
substantial information and does not justify the added computational cost.

\bibliographystyle{apalike}
\bibliography{CFU_count}

\end{document}